\documentclass[nofootinbib,twocolumn,aps,prd,amsmath,superscriptaddress]{revtex4-1}
\usepackage{setspace}
\usepackage{xcolor}
\usepackage{fancyhdr}
\usepackage{graphicx}
\usepackage[ansinew]{inputenc}
\usepackage{amssymb}
\usepackage{amsmath}
\usepackage{cancel}
\usepackage{float}
\usepackage{subfigure}
\usepackage{cancel}
\usepackage{tabularx}

\begin{document}
	\title{Which first order phase transitions to quark matter are possible in neutron stars?}
	\author{Jan-Erik Christian}
	\email{jan-erik.christian@uni-hamburg.de}
	\affiliation{Hamburger Sternwarte, University of Hamburg, Gojenbergsweg 112, 21029 Hamburg, Germany}
	\author{J\"urgen Schaffner-Bielich}
	\email{schaffner@astro.uni-frankfurt.de}
	\affiliation{Institut f\"ur Theoretische Physik, Goethe Universit\"at, Max-von-Laue-Stra\ss e 1, 
		D-60438 Frankfurt am Main, Germany}
	\author{Stephan Rosswog}
	\email{stephan.rosswog@uni-hamburg.de}
	\affiliation{Hamburger Sternwarte, University of Hamburg, Gojenbergsweg 112, 21029 Hamburg, Germany}
	\affiliation{The Oskar Klein Centre, Department of Astronomy, AlbaNova, Stockholm University, SE-106 91 Stockholm, Sweden}
	\date{\today}
	\begin{abstract}
		We examine which first order phase transitions are consistent with today's astrophysical constraints. In particular, we explore how a well-constrained mass-radius data point would restrict the admissible parameter space and to this end, we employ the most likely candidates of the recent NICER limits of PSR J0030+0451. To systematically vary the stiffness of the equation of state, we employ a parametrizable relativistic mean field equation of state, which is in compliance with results from chiral effective field theory. We model phase transitions via Maxwell constructions and parametrize them by means of the transitional pressure $p_{\rm trans}$ and the jump in energy density $\Delta\epsilon$. This provides us with a generic setup that allows for rather general conclusions to be drawn. We outline some regions in the $p_{\rm trans}$-$\Delta\epsilon$ parameter space that may allow for a phase transition identification in the near future. We also find that a strongly constrained data point, at either exceptionally large or small radii, would reduce the parameter space to such an extent that mass and radius become insufficient indicators of a phase transition.
	\end{abstract}
	\maketitle
	
	
	\section{Introduction}\label{incanus}
	The equation of state (EoS) of dense matter beyond $\sim 1.5$ times the nuclear saturation density is poorly known to date.
	Neutron stars are uniquely suited to shed light on the properties of this matter state that is not accessible to terrestrial experiments, and mass measurements have played a crucial role in that.
	With the increase of gravitational wave detection sensitivity the radius of neutron stars can be constrained by the measurement of the tidal deformability from neutron star mergers. In addition, the measurement of the x-ray thermal emission of neutron stars allows to limit the compactness, the ratio of mass over radius and thereby constrain the mass-radius relation. Recent advances in these astrophysical observations open the possibility to probe the equation of state of neutron star matter \cite{Abbott:2018exr,Fattoyev:2020cws,Ghosh:2021bvw,Huth:2021bsp} and to investigate the interior of neutron stars for a first-order phase transition \cite{Drago:2017bnf,Alvarez-Castillo:2018pve,Bauswein:2018bma,Montana:2018bkb,Blaschke:2020vuy,Bombaci:2020vgw,Christian:2020xwz,Landry:2022rxu,Zhang:2023wqj}.
	
	Strictly speaking, a first order phase transition cannot be ruled out by a measurement based on general relativity
	alone. The strong equivalence dictates that gravity cannot distinguish between compact stars which differ just in their compositions. It holds for general relativity (GR) and has been confirmed 
	for compact stars to high precision (i.e.\ at the level of $10^{-6}$) 
	by the long-term measurement of the pulsar PSR J0337+1715 in a triple system with two white dwarfs
	\cite{Archibald:2018oxs,Voisin:2020lqi}. This blindness of GR applies as well if there is a first order phase transition in the interior of neutron stars to a new phase, altering just the composition. So a determination of the mass-radius diagram of neutron stars will not allow pinning down the composition or changes in the composition in the interior of neutron stars unless they lead to a substantial modification of the global structure of the neutron star.
	Actually, the presence of at least one first-order phase transition is well established in neutron star matter. The crust-core transition at around half saturation density is of first order, as the lattice of the crust disappears in the core,
	see Refs.\ \cite{Baym:1971ax,Pethick:1994ge,Hebeler:2013nza}. That first order phase transition produces a small jump in the energy density, implying a small change of the equation of state. The crust-core phase transition appears around the minimum of the neutron star mass-radius relation, i.e.\ at a mass of about $0.1M_\odot$ with a radius of about 200km. There might be other first order phase transitions appearing in the core of neutron stars such that they do not induce a pronounced change in the equation of state and in the mass-radius relation. 
	This effect is well known as the 'masquerade effect', pointed out by Alford, Braby, Paris, and Reddy in 2005 \cite{Alford:2004pf} for the case of the phase transition to quark matter. Such a masquerade effect cannot be excluded at present for other new phases appearing in the core of neutron stars, be it a phase transition to 
	hyperons \cite{Chatterjee:2015pua,Tolos:2016hhl,Oertel:2016xsn,Li:2018jvz,Fortin:2020qin,Burgio:2021vgk,Ghosh:2022lam,Rather:2021azv,Li:2023owg,Kochankovski:2022rid,Blacker:2023opp,Kochankovski:2023trc},
	$\Delta$-baryons \cite{Glendenning:1984jr,Huber:1997mg,Drago:2014oja,Kolomeitsev:2016ptu,Motta:2019ywl,Dexheimer:2021sxs,Sedrakian:2022ata,Marquez:2022gmu,Rather:2023dom}
	kaon condensates
	\cite{Thorsson:1993bu,Maruyama:1994np,Fujii:1995rh,Glendenning:1997ak,Glendenning:1998zx,Pons:2000xf,Banik:2001yw,Banik:2002qu,Menezes:2005ic,Thapa:2021kfo,Ma:2022knr}, 
	color-superconducting quark matter \cite{Alford:2002rj,Grigorian:2003vi,Drago:2004vu,Ippolito:2007hn,Pagliara:2007ph,Baym:2017whm,Blaschke:2021poc,Shahrbaf:2021cjz,Ivanytskyi:2022bjc}, 
	or quarkyonic matter
	\cite{Fukushima:2015bda,McLerran:2018hbz,Zhao:2020dvu,Sen:2020qcd,Kumar:2023qyu,Pang:2023dqj}.
	A similar masquerade effect could be present for the recently discovered chiral-spin phase seen in lattice QCD calculations \cite{Glozman:2022lda,Philipsen:2022wjj}, which would involve the phase transition to the parity partner of the nucleon in neutron star matter, see e.g.\ \cite{Dexheimer:2007tn,Dexheimer:2008cv,Mukherjee:2017jzi,Marczenko:2018jui,Gao:2022klm,Minamikawa:2023eky,Kong:2023nue,Fraga:2023wtd}.
	
	The present investigation of a first-order phase transition 
	should be put in perspective in view of the discussion above. We study the observability of a first-order phase transition 
	from the mass-radius relation of neutron stars below. That implies to study a substantial jump in the energy density in the equation of state to a new phase taking place in the core of neutron stars so that there are measurable effects on the mass-radius relation. The only ingredient of the property of the new phase is the equation of state after that jump in energy density, which neither specifies the particle composition of the new phase nor the particle physics features of the new phase, which e.g.\ could consist of nucleonic particles in a new (chirally restored or chiral-spin) phase. 
	For convenience, we will label the matter before the jump in energy density as being the 'hadronic matter', the matter
	after the jump as 'quark matter' keeping in mind the discussion above. 
	In this spirit, the equation of state after the jump will be parametrized without resorting to a detailed microscopic model. Therefore, our results are quite generic for phase transitions.
	
	We want to point out, in addition that
	the phase transition could be constructed with the general Gibbs criterion for two chemical potentials, which results
	in a mixed phase in neutron star matter. The existence of a mixed phase hinges on the unknown surface tension between the two phases. If a mixed phase is present it will produce two kinks in the equation of state and smear out the jump in energy-density.
	Also, it could be that the transition is not a true phase transition but a rapid cross-over. If the rapid cross-over 
	is sufficiently 'fast' or if there are finite-size effects for the mixed phase it will produce similar mass-radius relations as for the case of a true phase transition \cite{Alvarez-Castillo:2013spa,Yasutake:2014oxa}. 
	Hence, our results should equally well hold for a sudden change in the equation of state without a true first order phase transition being present. So we are labeling the sudden change to a new form of matter in the core of neutron stars as 'phase transition' keeping in mind that it can also stand for a rapid cross-over.
	
	In the following, we will apply the well established mass constraint of about $2\,M_\odot$ \cite{Demorest:2010bx,Antoniadis:2013pzd,Fonseca:2016tux,Cromartie:2019kug,Romani:2022jhd}, the comparatively recent radius constraint of about 11km determined by NICER \cite{Miller:2021qha,Riley:2021pdl,Raaijmakers:2021uju} for J0740+6620 and the tidal deformability constraint of $\tilde\Lambda = 720$ from the GW170817 gravitational wave measurement \cite{TheLIGOScientific:2017qsa,Abbott:2018wiz,Abbott_2020} to our EoS. We use a relativistic mean field EoS \cite{PhysRev.98.783,Duerr56,Walecka74,Boguta:1977xi,Serot:1984ey,Mueller:1996pm,Typel:2009sy,Hornick:2018kfi} with variable effective masses, which allows us to investigate a large range of EoSs with varied stiffness. Further, a phase transition at a critical pressure to quark matter is added, resulting in stars with a hadronic mantle and a quark core, a so-called hybrid star \cite{Ivanenko:1965dg,Itoh:1970uw,Alford:2004pf,Coelho:2010fv,Chen:2011my,Masuda:2012kf,Yasutake:2014oxa,Zacchi:2015oma,Xie:2020rwg}. 
	The possibility of such a configuration changes the mass-radius relation in a substantial way from a purely hadronic approach. 
	However, not all parameter sets result in mass-radius
	relations that can be confidently detected with current facilities.
	Most of these parameter sets lead to so-called 'twin stars', where two stars have the same mass, but different radii \cite{Kampfer:1981yr,Glendenning:1998ag,Schertler:2000xq,SchaffnerBielich:2002ki,Zdunik:2012dj,Alford:2015dpa,Blaschke:2015uva,Zacchi:2016tjw,Alford:2017qgh,Christian:2017jni,Blaschke:2019tbh,Tan:2020ics,Christian:2021uhd}.
	
	We examine the parameter space that allows for the detection of hybrid stars under all relevant astrophysical constraints, as well as two hypothetical precise data points with large and small volumes. We find that a phase transition is a good approach to ease the tension between GW170817 and radius constraints from NICER, but it is growing less likely to be detectable with standard astrophysical observables.
	
	\section{Equation of State}\label{EoS}
	
	\subsection{Hadronic Equation of State}
	
	For the hadronic EoS, we follow the work of Hornick et al.\ \cite{Hornick:2018kfi}, where parameter sets of a relativistic mean field approach \cite{PhysRev.98.783,Duerr56,Walecka74,Boguta:1977xi,Serot:1984ey,Mueller:1996pm,Typel:2009sy,ToddRutel:2005fa,Chen:2014sca,Hornick:2018kfi} in compliance with predictions from chiral effective field theory \cite{Drischler:2016djf} are discussed. In this approach the stiffness of the EoS can be controlled via the effective nucleon mass at saturation density $m^*/m$, where smaller values correspond to stiffer EoSs \cite{Boguta:1981px}. If we fix the slope parameter $L = 60\,\mathrm{MeV/fm^3}$ and symmetry energy $J = 32\,\mathrm{MeV/fm^3}$ at saturation density the range of allowed $m^*/m$ is maximal \cite{Hornick:2018kfi}. These values are well within experimentally determined ranges \cite{Lattimer:2023rpe}. 
	For this work, we use EoSs with effective masses between $m^*/m = 0.55$ and $m^*/m = 0.75$. Since the effective mass is tied to the softness 
	\cite{Boguta:1981px}, the softest of these EoSs is generated for $m^*/m = 0.75$ and has a maximal mass barely above $2\,M_\odot$, failing to meet the current mass constraint of at least $2.09\,M_\odot$ \cite{Romani:2022jhd}. The stiffest EoS generates a maximal mass of nearly $2.8\,M_\odot$, but is only marginally within the radius range derived from the NICER observation of the pulsar J0740+6620, which is about 11\,km - 15\,km \cite{Riley:2021pdl,Miller:2021qha,Raaijmakers:2021uju}. Furthermore, it is outside the tidal deformability constraint from GW170817 \cite{Abbott:2018wiz}. Only EoSs with $0.65 \le m^*/m < 0.75$ are compatible with NICER, GW170817 and the mass constraints without a phase transition present. However, particularly the compatibility with GW170817 can be improved with the inclusion of a phase transition, 
	because the resulting hybrid stars are more compact than their purely hadronic counterparts \cite{Paschalidis:2017qmb,Alvarez-Castillo:2018pve,Christian:2018jyd,Montana:2018bkb,Sieniawska:2018zzj,Christian:2019qer,Tsaloukidis:2022rus,Landry:2022rxu}.
	
	\subsection{Phase Transition} 
	\begin{figure*}
		\centering				
		\includegraphics[width=17cm]{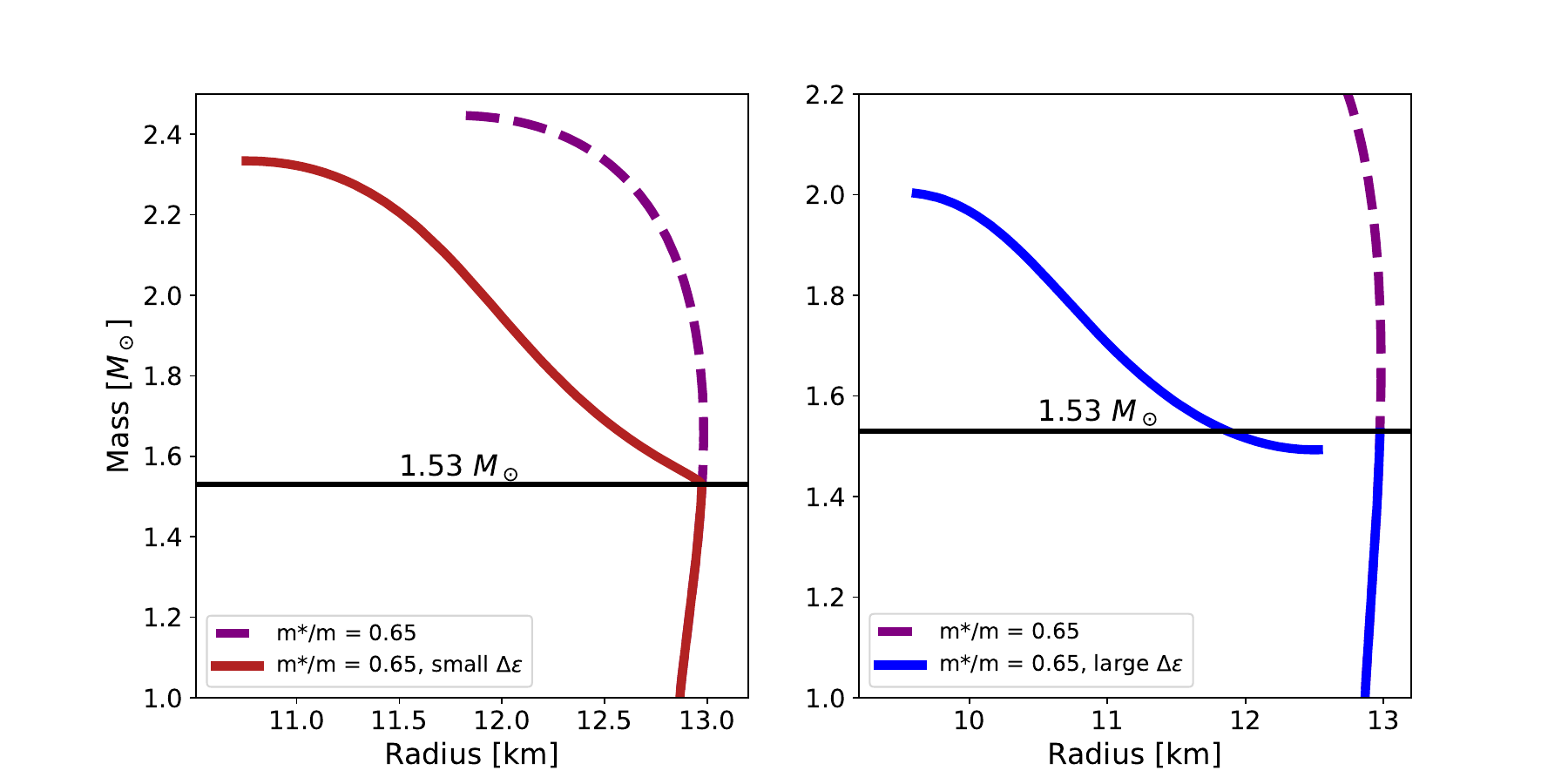}
		\caption{\footnotesize Schematic showcase of the change caused by small (left) or large (right) jumps in energy density. The pure hadronic mass-radius relation is shown as a purple dashed line. For both examples the phase transition to quark matter occurs at a pressure that corresponds to $1.53\,M_\odot$. For small $\Delta\epsilon$ the onset of the quark phase manifests as a kink in the mass-radius relation, for larger $\Delta\epsilon$ two separate branches are present.} 
		\label{twins}
	\end{figure*}

	We model the phase transition applying a Maxwell construction, where the energy density has a discontinuity of $\Delta\epsilon$ at the point of transition while the pressure $p = p_{\rm trans}$ and the chemical potential $\mu = \mu_{\rm trans}$ stay constant. The quark phase is described using the well established constant speed of sound approach \cite{Zdunik:2005kh,Alford:2013aca}. This means the EoS can be expressed as: 
	\begin{equation}
		\epsilon(p) =
		\begin{cases} 
			\epsilon_{HM}(p)	&  p < p_{\rm trans}\\
			\epsilon_{HM}(p_{\rm trans})+\Delta\epsilon + c_{QM}^{-2}(p-p_{\rm trans})	& p > p_{\rm trans}\\
		\end{cases}
	\end{equation}  
	where $c_{\rm QM}$ is the speed of sound of quark matter 
	(we use natural units and set $\hbar = c = 1$ in the following). We choose the maximally stiff quark phase by setting $c_{\rm QM} = 1$. This allows us to investigate the largest possible parameter space.

	\section{Twin stars}
	\begin{figure*}
		\centering				
		\includegraphics[width=19cm]{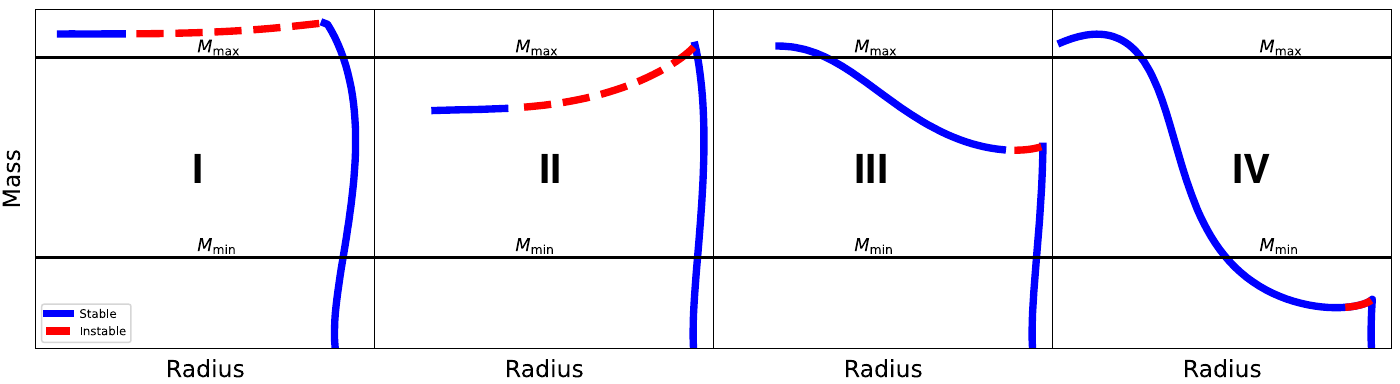}
		\caption{\footnotesize Sketch of representative examples of the (updated) four categories, similar to Fig.5 from \cite{Christian:2017jni}. $M_{\rm max}$ and $M_{\rm min}$ denote the highest and lowest known neutron star masses, respectively.}
		\label{Categories}
	\end{figure*}
	
	When the EoS contains a phase transition, as described above, the mass-radius sequence can become unstable close to the point of transition.
	Whether or not a star with a central pressure slightly above the pressure of a phase transition is stable is determined by $\Delta\epsilon$, where the exact value at which a central pressure of $p_{\rm cental} = p_{\rm trans}$ cannot generate a stable star is expressed as \cite{seidov:1971pty,Kaempfer:1981a}:  
	\begin{equation}\label{Eq:Seidov}
		\frac{\Delta\epsilon}{\epsilon_{\rm trans}} = \frac{1}{2} + \frac{3}{2}\frac{p_{\rm trans}}{\epsilon_{\rm trans}}\,,
	\end{equation}  
	which is sometimes referred to as Seidov limit. 
	Note that even $\Delta\epsilon$ values slightly below the Seidov limit will destabilize the mass-radius relation for 
	central pressures slightly above $p_{\rm trans}$, only $\Delta\epsilon$ far below the limit will not cause a 
	destabilization at all. This means, if the sequence becomes unstable, the maximum of the first branch will be located at a 
	similar mass and radius point, even for $\Delta\epsilon$ below the Seidov limit. 
	The sequence can regain stability if the central pressure increases rapidly enough with energy density.
	The position of this point is determined by 
	the jump in energy density of the phase transition as well. During the unstable segment of the mass-radius relation the 
	mass decreases as a function of central pressure \cite{Bardeen1966}, therefore the mass-radius relation has a local 
	minimum when the second stable branch starts. As a result, two stars with the same mass, but different radii can be generated by the same EoS. This configuration is called twin stars and can only be achieved with a substantial change in the EoS, 
	a first-order phase transition in the EoS or a rapid crossover,
	making twin stars ideal indicators for the presence of a new phase in the cores of neutron stars. To determine their usefulness it is helpful to sort twin star configurations into four categories \cite{Christian:2017jni}, see section \ref{section:cat}.
	
	We note that for most mass-radius relations, the second branch is partially or even fully located at smaller (baryonic) masses than the maximum of the first branch. For such cases the most obvious formation channel, accretion onto a neutron star, is not possible.
	Alternative scenarios include neutron star mergers \cite{Most:2018eaw,Bauswein:2018bma,Weih:2019prev} or core collapse supernovae \cite{Sagert:2008ka,Fischer:2010wp,Hempel:2015vlg,Fischer:2017lag}. Even, the accretion-induced collapse of a white dwarf could provide a potential path to such hybrid stars, especially if the central pressure is increased by the presence of dark matter, as suggested by \cite{Leung:2019ctw}, but this scenario has not been studied yet.
	
	In any case, if the jump in energy density is very small compared to the Seidov limit, the mass-radius relationship will not feature an unstable segment at all. Instead, the hadronic branch and the hybrid star branch are connected. While this means that it becomes impossible to find true twin stars, the mass-radius relation is still significantly altered by the presence of quark matter, with a noticeable kink at the point of transition.
	
	Exemplary mass-radius relations for small (left) and large (right) values of $\Delta\epsilon$ are shown in 
	Fig.~\ref{twins}. The mass at the point of transition is indicated by a black line, highlighting the presence of two stars with exactly the same mass, but different radii in the right plot and its absence in the left. The mass-radius relations are compared to the one generated by the original EoS without a phase transition, represented by a purple dashed line, where even for the connected case a clear deviation from the purely hadronic case is visible.
	In previous works it has been shown that the value of $\Delta\epsilon$ determines the position of the hybrid star branch in the mass-radius diagram, where decreasing the value of $\Delta\epsilon$ moves the hybrid branch closer to the maximum of the hadronic branch until the two branches connect for small $\Delta\epsilon$ \cite{Christian:2017jni}. The shape of the second branch is only determined by $p_{\rm trans}$ and the quark matter EoS.
	
	\section{Categorization}\label{section:cat}
	
	In previous works, we categorized twin star configurations into four distinct categories based on the position of their hadronic and hybrid star maxima \cite{Christian:2017jni}. In this work, we slightly adjust this definition by using the known mass ranges of neutron stars as a basis. This new categorization will therefore be more broadly applicable. The mass of the heaviest known neutron star will be referred to as $M_{\rm max}$. Likewise, the least massive known neutron star mass will be referred to as $M_{\rm min}$. Currently $M_{\rm max}$ is known to be about $2\,M_\odot$ 
	\cite{Demorest:2010bx,Antoniadis:2013pzd,Cromartie:2019kug,Romani:2022jhd} 
	and $M_{\rm min}\simeq1.174\pm 0.004\,M_\odot$ for the pulsar J0453+1559 \cite{Martinez:2015mya}
	(however, see \cite{Tauris:2019sho} for a different interpretation).
	
	This categorization presupposes that the mass-radius relation becomes unstable close to the point of transition. 
	However, since the shape of the second branch is determined by $p_{\rm trans}$ and the quark matter EoS, the trends outlined for these categories remain robust when the branches are connected. Even if the two branches are connected, it is still possible to find two stars with similar masses and significantly different radii and to estimate the transitional pressure and jump in energy density from their observation. Representative examples of the four categories are shown in Fig. \ref{Categories}.
	The four categories are the following:\\\\
	\textbf{Category I:} Both the hadronic and the hybrid star maximum are more massive than $M_{\rm max}$. This requires a large transitional pressure, which leads to a characteristic flat second branch, featuring large radius differences at masses close to $M_{\rm max}$.\\
	\textbf{Category II:} Only the pure hadronic neutron stars reach $M_{\rm max}$. Like category I, this necessitates a large value of $p_{\rm trans}$, causing a flat hybrid star branch. However, due to the increase in $\Delta\epsilon$ the second branch has smaller masses compared to category I. As a result, no hybrid stars of this category can be generated via mass accretion onto a neutron star.\\
	\textbf{Category III:} The hadronic maximum of this category is located above $M_{\rm min}$ and below $M_{\rm max}$, therefore requiring smaller $p_{\rm trans}$ compared to category I and II. This means this category is defined by the necessity to reach $M_{\rm max}$ in the hybrid star branch, putting an upper limit on $\Delta\epsilon$. Due to the smaller $p_{\rm trans}$ the hybrid star branch in this category is much steeper than in the previous two categories, decreasing the radius differences between twin stars.\\
	\textbf{Category IV:} This category is defined by its early phase transition. The maximal mass of the first branch is below $M_{\rm min}$, meaning only hybrid stars would be observable. This makes it impossible to use the differences in radius as an indicator for a phase transition.
	
	\section{Parameter space constraints}
	\begin{figure*}
		\centering				
		\includegraphics[width=18cm]{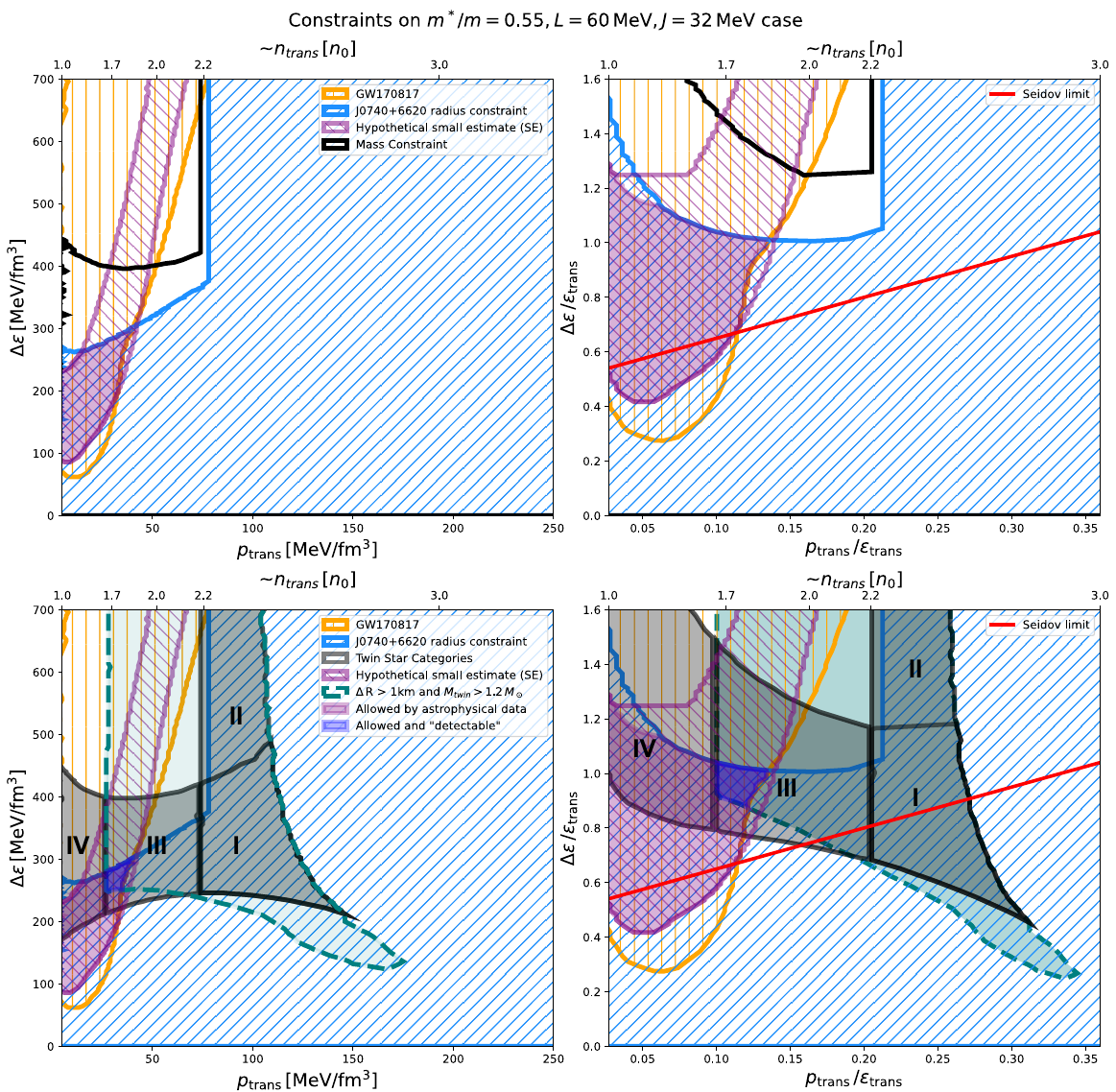}
		\caption{\footnotesize Parameter space for phase transitions following an EoS with $m^{*}/m = 0.55$. On the left side absolute values are shown, while on the right side the values are divided by the energy density in the hadronic phase at the point of transition, for ease of comparison. The x-axis shows the transitional pressure, as well as the corresponding baryon density. On the y-axis values of $\Delta\epsilon$ are shown. In the upper row only the astrophysical constraints (including the hypothetical one) on the parameter space are shown. In this row all parameter sets below and to the right of the black line meet the mass constraint. This line is identical with the upper bound of the twin star area in the second row. The hatched light blue area indicates all parameter sets that generate results compatible with the NICER measurement of J0740+6620. The orange hatched area indicates the constraint from the tidal deformability determined in GW170817. The purple hatched area is the constraint a hypothetical small neutron star would impose, based on the NICER estimate of J0030+0451. The region where all these astrophysical constraints overlap is shaded purple. In the second row we highlight the areas where mass and radius measurements are particularly promising indicators of a phase transition. The black shaded area indicates where twin stars are possible and is further divided into the four categories. The teal shaded area indicates parameter sets generating two stars with similar masses but radius differences larger than 1\,km. For this EoS GW170817 is the strongest constraint and requires small values of $p_{\rm trans}$, which align very well with the hypothetical constraint from J0030+0451. Only a small area (shaded blue) is both possible and reasonably well testable with mass and radius measurements.}
		\label{M55}
	\end{figure*}
	
	\begin{figure*}
		\centering				
		\includegraphics[width=18cm]{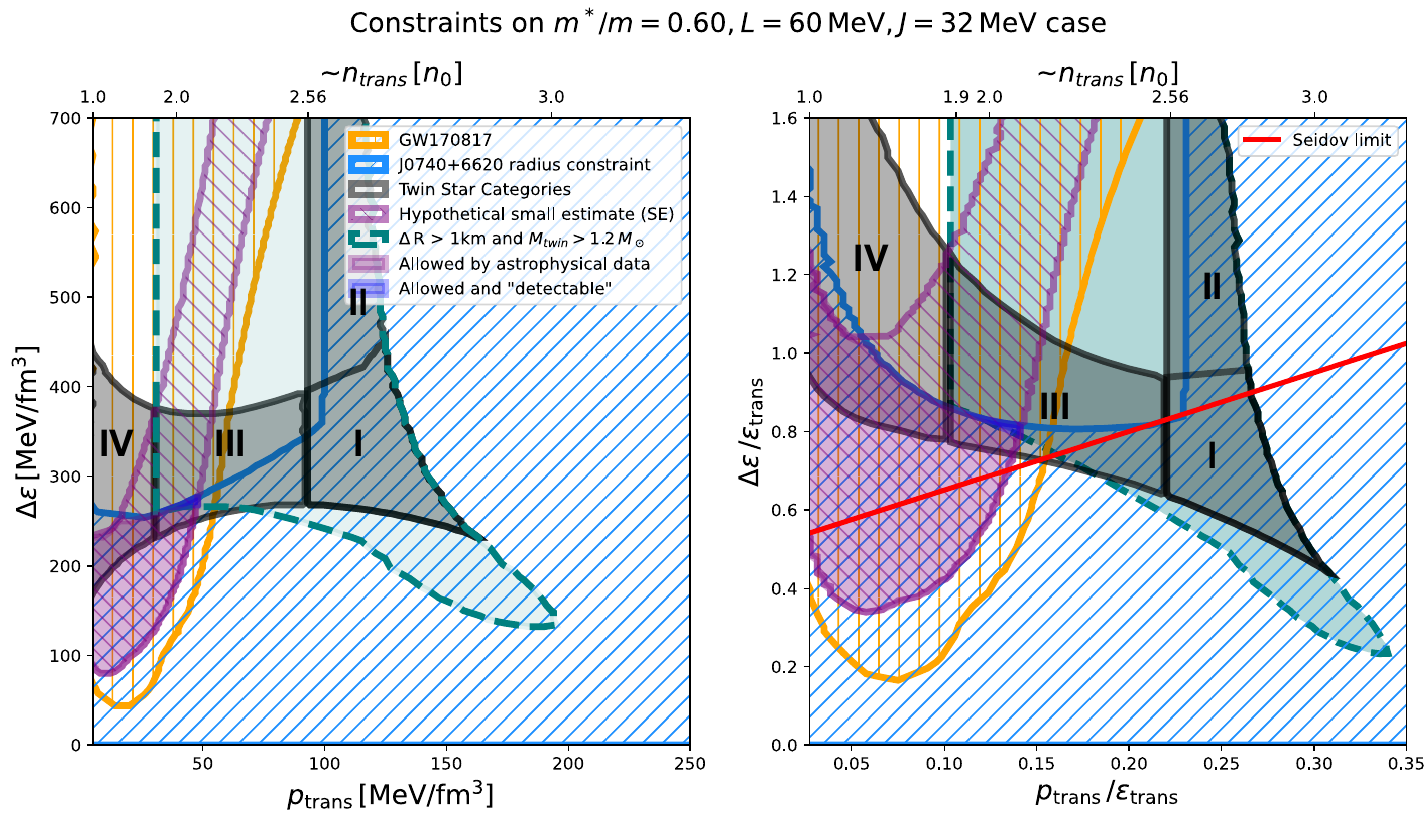}
		\caption{\footnotesize Parameter space for phase transitions following an EoS with $m^{*}/m = 0.60$. GW170817 is the strongest constraint. A hypothetical small neutron star measurement from NICER would remove the otherwise still accessible area, where detections using mass-radius measurements would indicate a phase transition from the parameter space.}
		\label{M60}
	\end{figure*}
	
	\begin{figure*}
		\centering				
		\includegraphics[width=18cm]{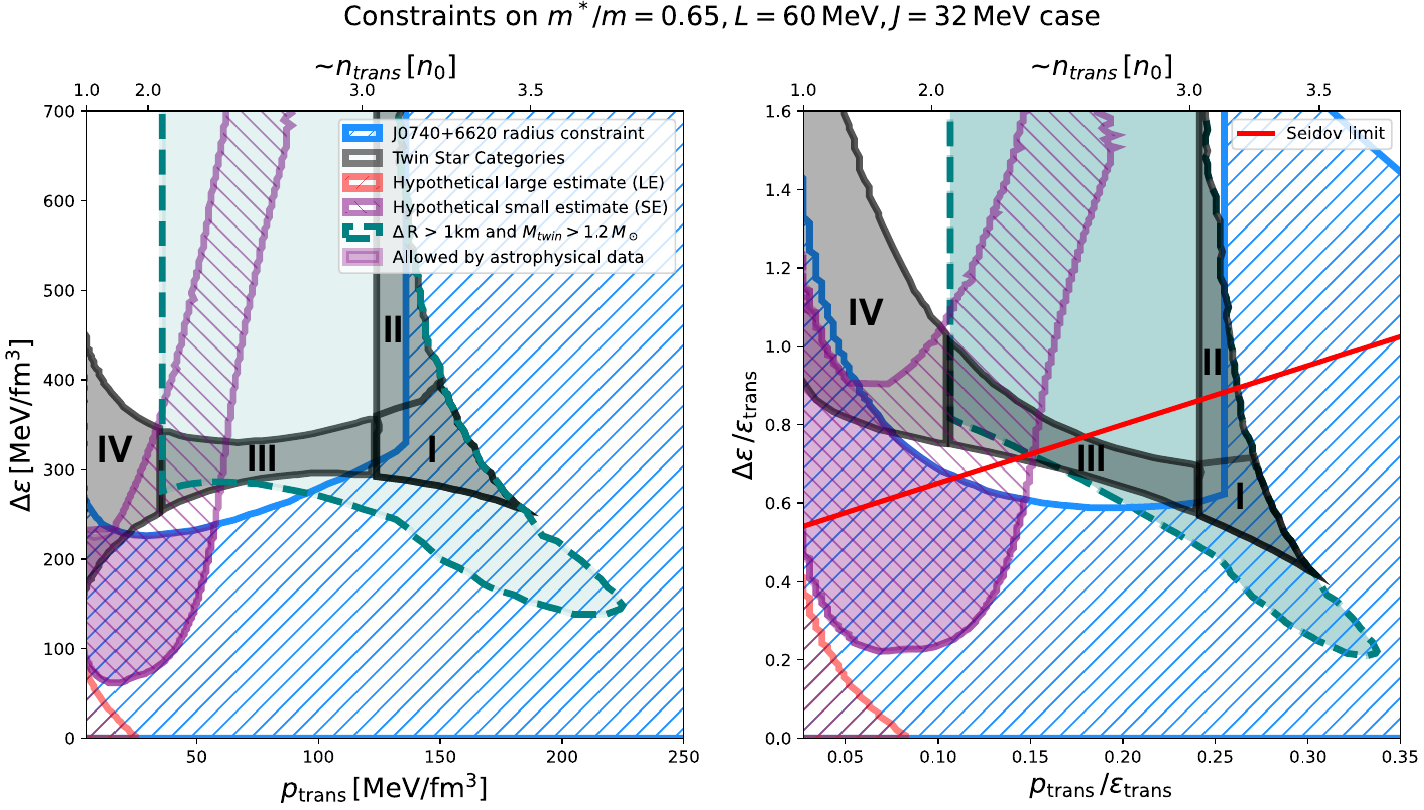}
		\caption{\footnotesize Parameter space for phase transitions following an EoS with $m^{*}/m = 0.65$. The NICER constraint (hatched light blue area) from J0740+6620 reduces the allowed parameter space significantly. For large transitional pressures the allowed parameter sets remain easily detectable using mass-radius measurements. However, the small estimate of J0030+0451 would require small values of $p_{\rm trans}$, making it difficult to find a clear signal for a phase transition. The entire parameter space is able to reproduce the tidal deformability required by GW170817. A hypothetical large compact star estimate is added as a red hatched area. Such an object would only be realizable for extremely small transitional pressures and jumps in energy density.}
		\label{M65}
	\end{figure*}
	
	\begin{figure*}
		\centering				
		\includegraphics[width=18cm]{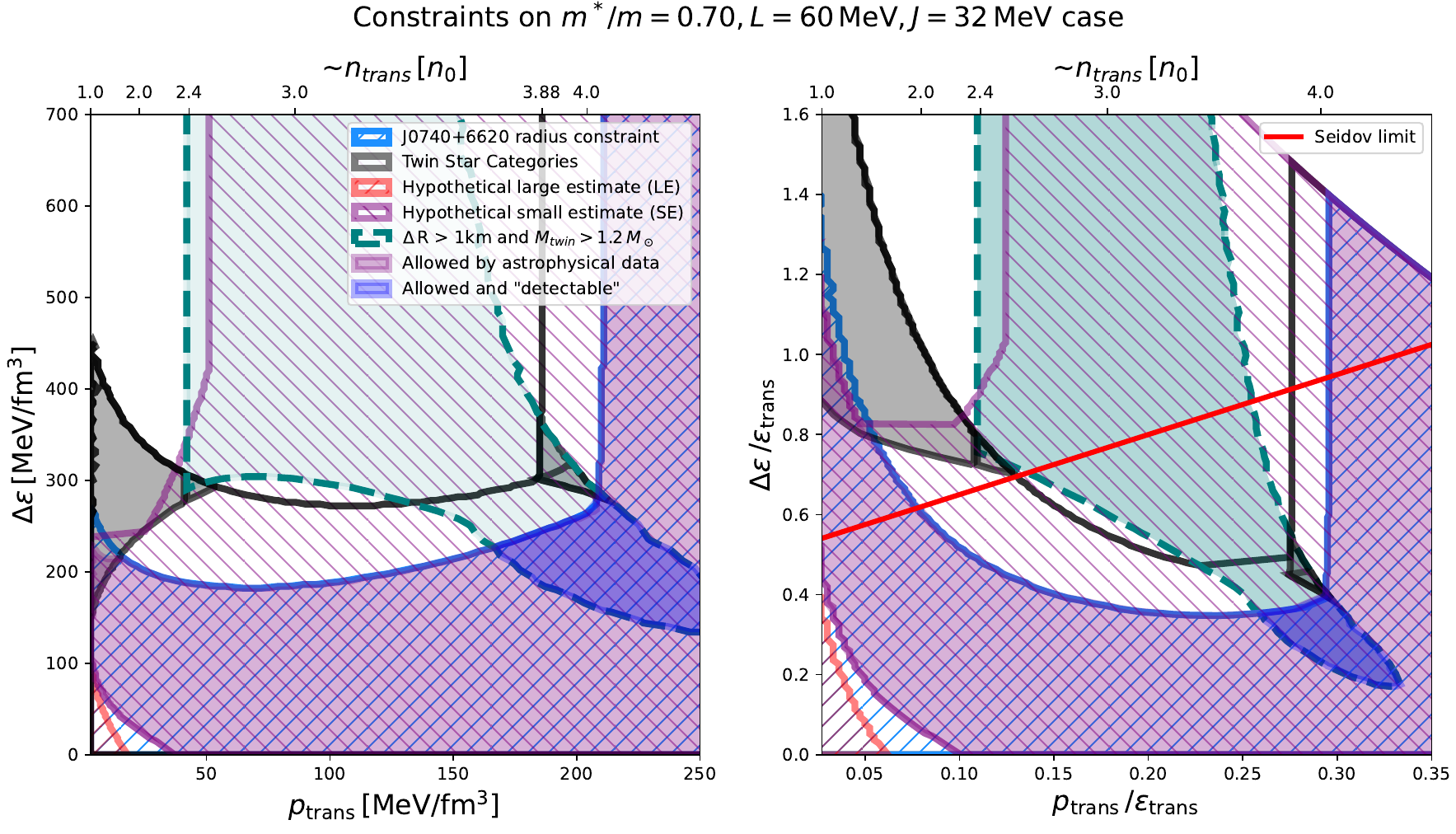}
		\caption{\footnotesize  Parameter space for phase transitions following an EoS with $m^{*}/m = 0.70$. The NICER constraint for J0740+6620 rules out category III and IV. Category I can be partially realized, but is relegated to a small area in the parameter space due to the softness of the EoS. Unlike for the $m^{*}/m = 0.65$ it is possible to find similarly massive stars with over 1\,km radius difference at high transitional pressures. A hypothetical well constrained small estimate of J0030+0451 would not constrain the parameter space to a significant degree, while the large estimate would require small $p_{\rm trans}$ and $\Delta\epsilon$. }
		\label{M70}
	\end{figure*}
	
	\begin{figure*}
		\centering				
		\includegraphics[width=18cm]{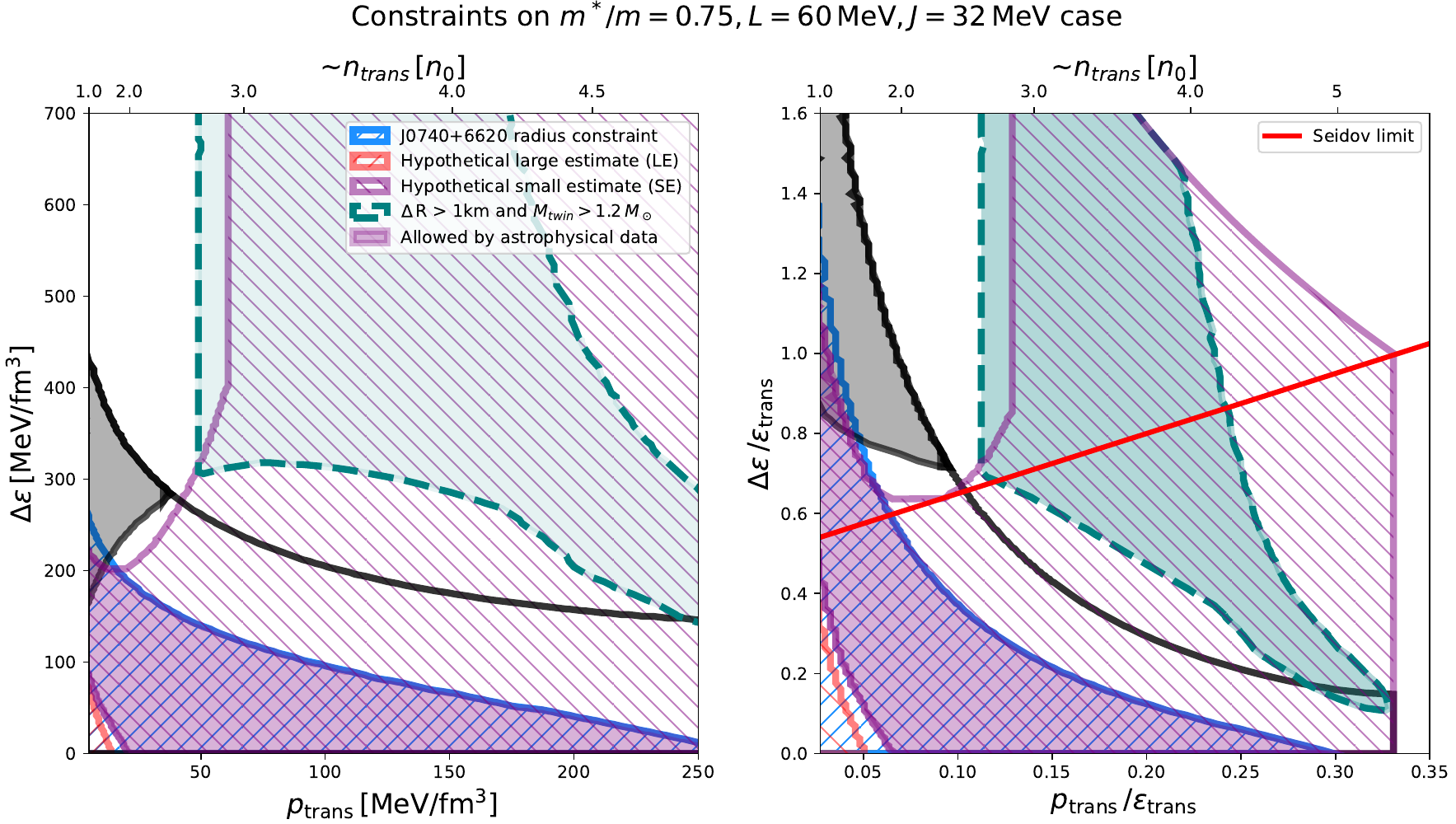}
		\caption{\footnotesize  Parameter space for phase transitions following an EoS with $m^{*}/m = 0.75$. The radius constraint from J0740+6620 manifests as a comparatively small upper limit for $\Delta\epsilon$. In the mass-radius relation this is visible as a backward kink to larger radii at the point of transition. No other astrophysical observables put constraints on this EoS.
		}
		\label{M75}
	\end{figure*}
	
	In this section we examine the constraints imposed on the equations of state discussed in section \ref{EoS} by the currently established astrophysical data points.
	
	Without a quark phase present, only the cases $m^*/m = 0.65$ and $m^*/m = 0.70$ are compatible with mass constraints \cite{Demorest:2010bx,Antoniadis:2013pzd,Fonseca:2016tux,Cromartie:2019kug,Romani:2022jhd}, NICER constraints \cite{Miller:2019cac,Riley:2019yda,Raaijmakers:2019qny,Miller:2021qha,Riley:2021pdl,Raaijmakers:2021uju} and the GW170817 tidal deformability constraint \cite{TheLIGOScientific:2017qsa,Abbott:2018wiz}. As we will show, with a phase transition both softer and stiffer equations of state can be adjusted to fit the available data, but only in a small part of the parameter space.
	
	Stiff EoSs are constrained strongly by GW170817 and soft EoSs are constrained by the NICER radius constraint. In Fig. \ref{M55} to \ref{M75} the parameter spaces for varied effective masses are shown, where the areas containing parameter sets that comply with a specific constraint are highlighted by color. 
	The sets resulting in the four categories are shaded black. They are ordered as follows: Category IV is on the left, Category III in the middle, Category II in the upper right and Category I in the lower right corner. The upper limit of the categories coincides with the mass constraint, since $M_{\rm data}$ defines the categories. For softer equations of state the parameter space where twin stars are realized shrinks, in these cases the mass constraint is shown with a black line, with parameter sets below that line 
	fulfilling the constraint.
	
	In addition to the categories we highlight the parameter sets which result in stars with nearly identical masses, i.e. $\Delta M \le 0.05\,M_\odot$, but significantly different radii ($\Delta R\ge\,$1km). We argue that such configurations have an enhanced chance to be detected by current facilities.
	
	The absolute values are shown on the left side of these plots, the relative values on the right. The relative values are advantageous when comparing the different EoSs, since a position in the diagram can be associated with a mass-radius shape more easily than with absolute numbers; however, they are less intuitive. Their advantage is demonstrated by the fact that, for all EoSs, the categories in the right hand plot are at the same position, only reduced to a smaller area with increased effective masses.
	To further exemplify the construction of these figures we split Fig. \ref{M55} into two rows. In the first row only the astrophysical constraints are shown, including a hypothetical constraint discussed later in this section. In the second row the areas including noticeable mass-radius indicators for a phase transition are added.
	
	The stiffest EoS (Fig. \ref{M55}; $m^\ast/m=0.55$) is mostly constrained by GW170817, which results in a strict upper limit for the transition pressure. A small value of $p_{\rm trans}$ is required, leading to category I and II being ruled out completely, but even category III is only realizable in half its parameter range. The jump in energy density is nearly unconstrained by the constraint from the tidal deformability, but the lower radius limit of about 11\,km from NICER \cite{Miller:2021qha,Riley:2021pdl,Raaijmakers:2021uju} restricts the allowed area to comparatively small jumps, cutting category IV and III in half in the $\Delta\epsilon$ range as well. This is because an increased jump in energy density moves the hybrid branch to more compact configurations with smaller radii, an effect that is enhanced by the early phase transition.
	For this EoS there is a small area where all constraints overlap and the mass-radius diagrams contain twin stars that feature a radius difference of $\Delta R \ge\,$1km. In this small area mass-radius measurements would be able to signal a phase transition (or a rapid crossover). 
	
	In a previous study we ruled out the presence of twin stars, if the phase transition takes place at densities below about $1.7\,n_0$ \cite{Christian:2020xwz}. We argued that twin stars require 'strong' phase transitions, which we defined as $\Delta\epsilon \ge 350\,\mathrm{MeV/fm^3}$. Such large jumps would move the hybrid star branch to radii below the NICER constraint. However, this argument would not rule out weaker transitions, where the branches are connected, or the distance between the branches is small.
	The present work reinforces our finding, based on a different argument. In Fig. \ref{M55} the border between category III and IV corresponds to $n \simeq 1.7\,n_0$. Since Fig. \ref{M55} is the stiffest case, this is the smallest density where the border between these two categories can lie. A category IV parameter set implies that the transition takes place at such small densities that only hybrid stars are realized in nature. As a result we would not expect to detect twin stars at all, if the phase transition takes place at densities smaller than $n \simeq 1.7\,n_0$, independent from the 'strength' of the transition.
	
	The second stiffest EoS used here (Fig. \ref{M60}; $m^\ast/m= 0.60$) is similarly constrained by GW170817 to small values of $p_{\rm trans}$. However, the radius constraint from NICER has a bigger impact on the allowed area than in the stiffer case, because the mass-radius relation is already located at smaller radii. This means, that not only category I and II are ruled out, but category III is nearly ruled out as well. Twin stars are still possible, but the radius differences drop below 1km making them unlikely to be detectable.
	
	Fig. \ref{M65} ($m^\ast/m=0.65$) shows the results for the middle value in this study. 
	We point out that while fits to properties of nuclei within the standard RMF model prefer effective masses in the range of $m^*/m=0.55$ to $0.60$, extended relativistic models with a tensor term are compatible with properties of nuclei 
	for an effective mass of up to $m^*/m=0.70$ \cite{Rufa:1988zz,Furnstahl:1997tk}
	and for $m^*/m = 0.67$ within the relativistic density functional model with tensor terms 
	(see \cite{Typel:2020ozc} and discussions therein).
	Since this EoS is compatible with the tidal deformability constraint without a phase transition, there is no limit on the transitional pressure, it is always possible to find such a value of $\Delta\epsilon$ that the parameter set is in agreement with all constraints. As this EoS is softer than the previous one, the NICER constraint results in even smaller allowed jumps in energy density. This rules out category III solutions completely, but not category I and II. These latter two categories include particularly large radius differences between similarly massive stars, due to their characteristic flat branches (for a detailed analysis see\cite{Christian:2021uhd}). Furthermore, since large values of $p_{\rm trans}$ are possible in this setup, the constraint from NICER can be met in the hadronic branch, leaving $\Delta\epsilon$ unconstrained following that point. This is visible in the plot as a vertical line in the NICER constraint.
	
	As the EoSs get softer, it becomes more difficult to produce the four categories. This is evident when comparing Fig.~\ref{M65} with Fig.~\ref{M70}. In the softer $m^*/m = 0.70$ case all categories are significantly reduced in parameter space size. This is because high central pressures increase the difficulty of regaining stability and the same central pressure in a soft EoS generates a less massive star than in a harder case. For the $m^*/m = 0.70$ case, this means nearly all categories are ruled out, except for small segments of category I and IV. However, it is still possible to generate large differences in radius for similar masses, while meeting all constraints. Such configurations are similar to category I solutions, but the branches are connected.
	
	Finally, the softest case (Fig. \ref{M75}; $m^\ast/m=0.75$) is only compatible with category IV and is heavily constrained by NICER to very small values of $\Delta\epsilon$, making nearly all possible configurations connected cases that would be indistinguishable from an EoS without a phase transition.
	
\begin{figure*}
		\centering				
		\includegraphics[width=18cm]{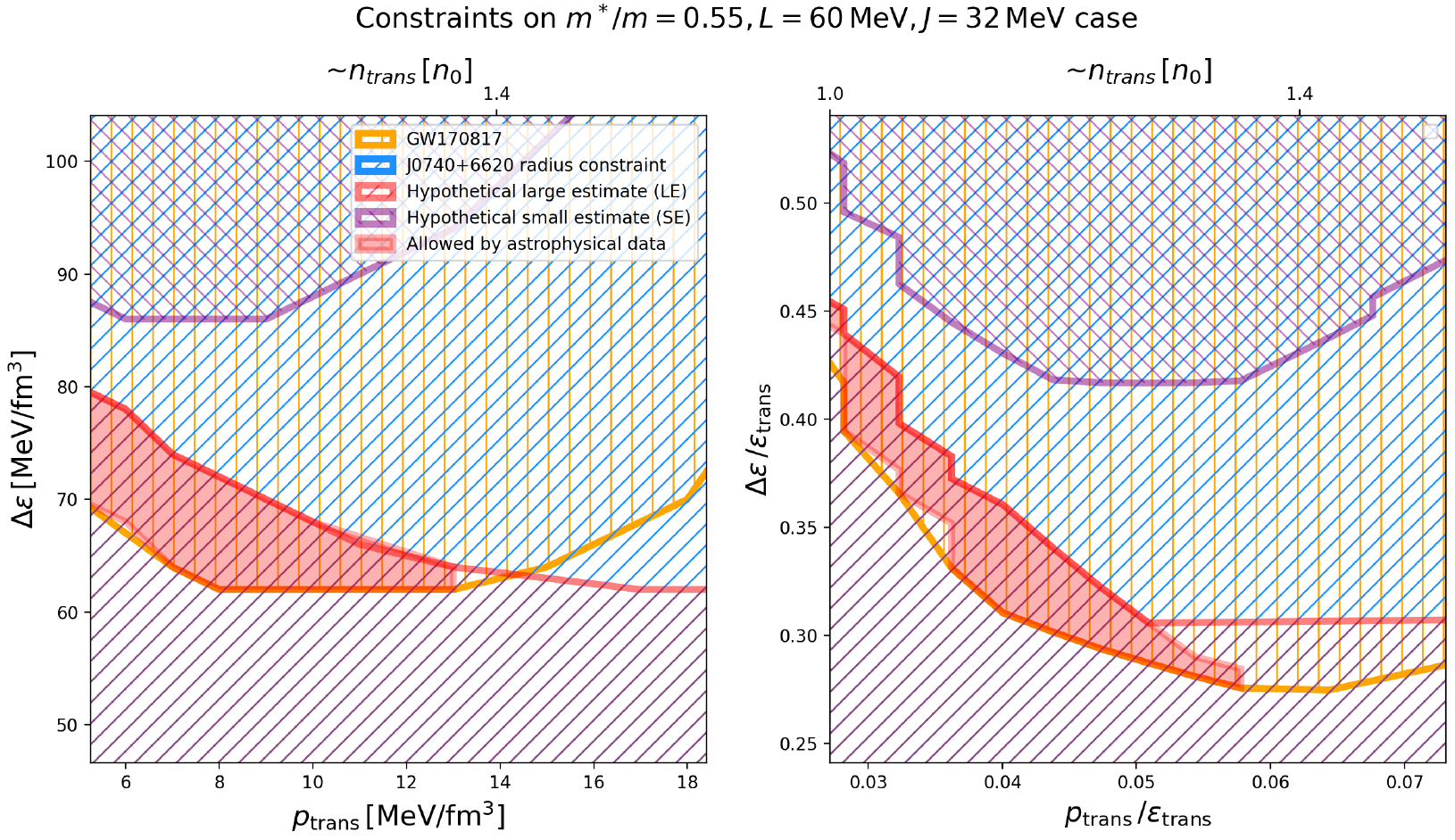}
		\caption{\footnotesize The overlap of GW170817 (orange) and the hypothetical larger estimate (red) contains the only parameter sets possible for the $m^*/m = 0.55$ case. It is noticeably small and quark matter dominated.}
		\label{LCE55}
	\end{figure*}

\begin{figure*}
		\centering				
		\includegraphics[width=18cm]{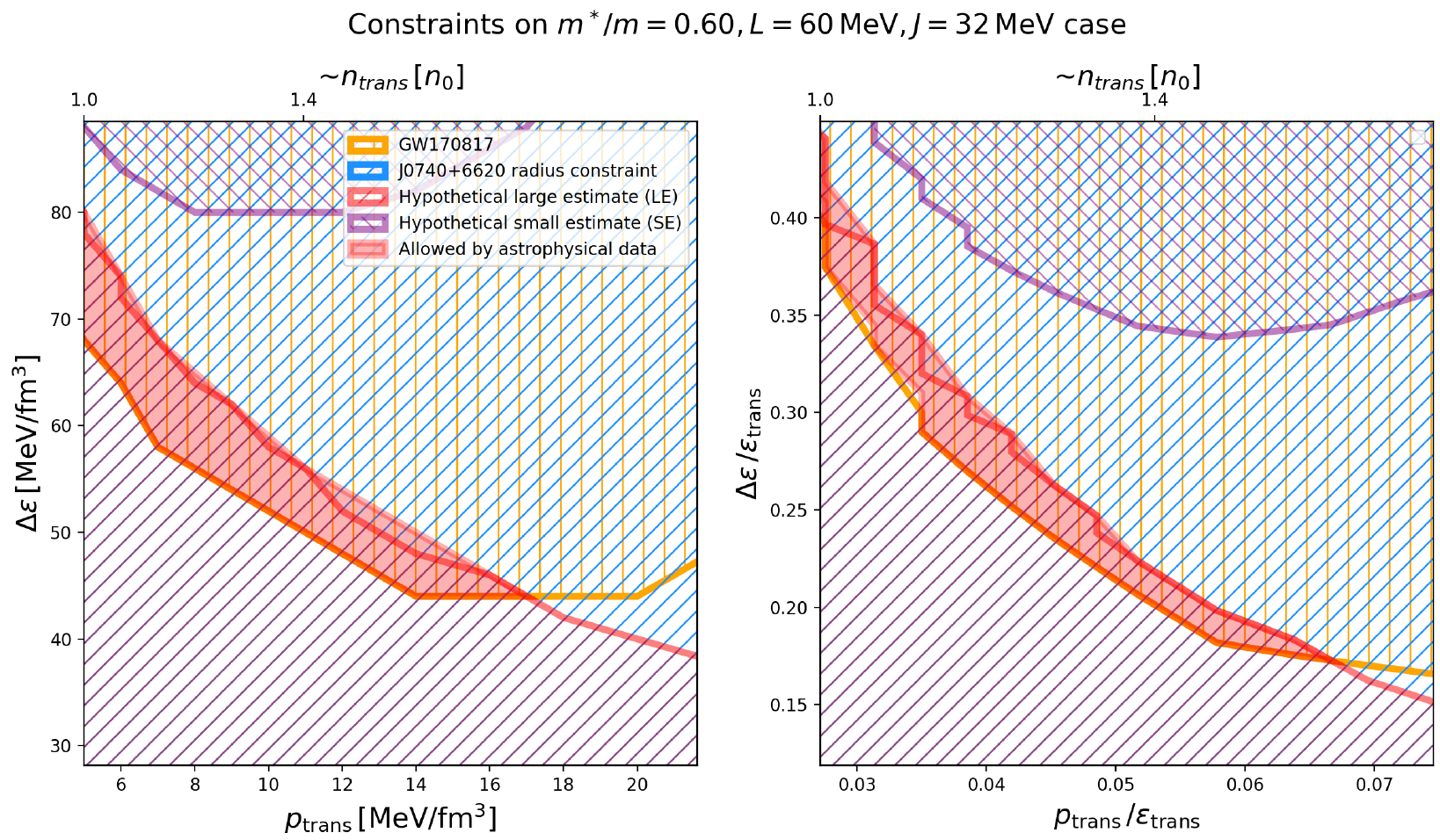}
		\caption{\footnotesize  The overlap of GW170817 (orange) and the hypothetical large estimate (red) contains the only possible parameter sets for the $m^*/m = 0.60$ case, should a LE like star be measured. It is even smaller than the $m^*/m = 0.55$ case and likewise quark matter dominated.} 
		\label{LCE60}
	\end{figure*}

	Vinciguerra et al. \cite{Vinciguerra:2023qxq} reanalyzed in a recent work the data from the first NICER radius measurement of PSR J0030+0451 and found it to be consistent with the previous estimates \cite{Riley:2019yda,Miller:2019cac,Raaijmakers:2019qny}. However, in addition they found a set of more complex models that also fit the data. Depending on the model chosen, the constraints on the mass and radius are much more stringent than otherwise, where they suggest, that PSR J0030+0451 has either a mass of $1.7^{+0.18}_{-0.19}\,M_\odot$ with a radius of $14.44^{+0.88}_{-1.05}\,$km, or a mass of $1.4^{+0.13}_{-0.12}\,M_\odot$ and a radius of $11.71^{+0.88}_{-0.83}\,$km. In the following, we assume that one of these data points is correct and analyze how severe the constraints on an equation of state with a phase transition would be, starting with the larger estimate (LE). A well established measurement of a neutron star with $1.7\,M_\odot$ at about 14.5\,km would rule out all equations of state in our approach that do not feature a phase transition, because only the stiff EoSs ($m^*/m \le 0.60$) are within the margin or error. Furthermore, a LE like neutron star would rule out virtually all sets of transition parameters for these stiff EoSs, because GW170817 requires a small transitional pressure, shifting the results to more compact configurations and outside the hypothetical LE range. This would constrain the allowed parameter sets to a small quark matter dominated area, which is shown in Fig. \ref{LCE55} and Fig. \ref{LCE60}.
	
	The parameter space for softer equations of state that can be accommodated by a LE like data point is slightly larger than for the stiffer EoSs, but still very small. The corresponding regions in the parameter sets are highlighted in red in Figs. \ref{M65} to \ref{M75}. Very small values of $p_{\rm trans}$, as well as a small jump in $\Delta\epsilon$ are required, heavily reducing the chance of detecting the presence of a phase transition in the mass-radius relation.
	
	Even though the parameter space for all cases is exceedingly small, one could argue that the detection of a LE like neutron star would point toward a phase transition being present, based on the severe tension that such a data point would have with the well established tidal deformability constraint, which cannot be mended (in this model) without a phase transition.
	
	If we assume that the smaller estimate (SE) is the correct one, the $m^*/m = 0.65$ case without a phase transition has to be ruled out. This means only the $m^*/m = 0.70$ case would fit with all astrophysical constraints, if no phase transition is present. However, the constraints put on the EoSs with a phase transition by SE would be less severe than by LE. In Fig. \ref{M55} to \ref{M75} the area populated by parameter sets in agreement with a potential SE measurement is shaded in purple. It is only a less severe constraint than LE, because it constrains largely the same parameter space as GW170817, especially for the stiff EoSs. The only case with a significantly reduced parameter space is the $m^*/m = 0.65$ case. With a SE like star all categories would be ruled out and no large radius differences for similarly massive stars would be expected for this hadronic EoS. The phase transition would have to take place at small transitional pressures, so the SE constraint can be satisfied in the hybrid star branch. It is unsurprising, yet worth noting, that SE and LE are mutually exclusive. There is no equation of state that could contain both stars, even with a phase transition.

	\section{Conclusion}\label{conclusion}
	
	Based on currently known astrophysical constraints, we analyzed which parts of the parameter space could be explained by first order phase transitions from a RMF EoS to quark matter.
	Our RMF EoS is parameterized in such a way that the stiffness can be adjusted by varying the effective nucleon mass. 
	This allows us to control the stiffness of the EoS in the high energy density regime, whereas changes in the slope parameter $L$ or symmetry energy $J$ would change the EoS behavior at smaller densities \cite{Hornick:2018kfi,Ghosh:2021bvw,Ghosh:2022lam}.
	However, it stands to reason that other ways of varying the stiffness would yield similar results.
	
	Using the established astrophysical constraints of mass, radius and tidal deformability, we find that only a very limited range of EoSs with moderate stiffness can accommodate all the data, in our approach $m^*/m = 0.65$ and $m^*/m = 0.70$. The range of viable hadronic EoSs increases if a phase transition is introduced. This is particularly the case for stiff equations of state. By themselves, those EoSs are incompatible with the upper limit on the tidal deformability, but phase transitions to hybrid stars could make them compact enough to be in agreement. However, the parameter space where these stiff EoSs are possible is small and requires a transition at comparatively small pressures. As a result, it would be hard to distinguish stiff hadronic EoSs with a phase transition from more exotic purely hadronic approaches.
	
	Softer EoSs already meet the tidal deformability constraint, instead their strongest constraint comes from the NICER radius measurement of J0740+6620. For such cases, the parameter space is reduced to two areas. Either the transition takes place at large central pressures, where all astrophysical constraints can be met by pure hadronic stars, or the transition takes place at very small pressure values, where the resulting mass-radius relation again would be hard to differentiate from an exotic purely hadronic case.
	
	In summary, the tidal deformability constraint determines the parameter space for stiff equations of state, whereas the parameter space for soft EoSs is determined by the radius constraint.
	
	We further considered a hypothetical more stringent radius constraint based on the recent work of Vinciguerra et al. \cite{Vinciguerra:2023qxq}, where the NICER data of the pulsar J0030+0451 was reanalyzed. Treating the largest neutron star model suggested by Vinciguerra et al., which has a mass of $1.7^{+0.18}_{-0.19}\,M_\odot$ and a radius of $14.44^{+0.88}_{-1.05}\,$km, as a concrete measurement, we would find that there is no hadronic equation of state that is compatible with all astrophysical constraints. However, it is still possible to construct a phase transition taking place in a stiff equation of state that meets the tidal deformability constraint and the constraint from the large hypothetical estimate. Such a phase transition requires small transitional pressures and would not result in any noticeable discontinuities in the mass-radius relation. Nevertheless, the detection of a star like this would heighten the tension between tidal deformability constraint and radius constraint to a point where a phase transition or other exotic model, such as the similar two family scenario \cite{Drago:2015cea,Drago:2015dea,Drago:2017bnf} is required to solve it.
	
	If we treat the smallest model of $1.4^{+0.13}_{-0.12}\,M_\odot$ with a radius of $11.71^{+0.88}_{-0.83}\,$km as a concrete measurement instead, no additional constraints can be put on the purely hadronic equations of state. This is also the case for stiff equations of state with a phase transition, where this hypothetical small neutron star data point occupies a similar region in the parameter space as GW170817. For our approach, only the model with the median effective mass of $m^*/m = 0.65$ would be constrained additionally to a noticeably degree by such a small neutron star. The resulting mass-radius relations would not feature large differences in radii at similar masses, making it difficult to differentiate a potential phase transition from purely hadronic cases.
	
	We find that current astrophysical measurements are getting stringent enough to put serious constraints on the equation of state. This means it becomes increasingly unlikely to be able to infer the presence of a phase transition from mass-radius relations alone. This problem would be compounded by a well constrained mass-radius data point at either particularly small or large stellar volumes. 
	However, differences in tidal deformability could still point towards a phase transition, once more data is available \cite{Landry:2022rxu}. 
	
	\begin{acknowledgments}
		JEC and SR were funded by the European Research Council (ERC) Advanced 
		Grant INSPIRATION under the European Union's Horizon 2020 research 
		and innovation programme (Grant agreement No. 101053985). SR was further supported by the Swedish Research Council (VR) under 
		grant number 2020-05044, by the research environment grant
		``Gravitational Radiation and Electromagnetic Astrophysical
		Transients'' (GREAT) funded by the Swedish Research Council (VR) 
		under Dnr 2016-06012, by the Knut and Alice Wallenberg Foundation
		under grant Dnr. KAW 2019.0112 and by the Deutsche 
		Forschungsgemeinschaft (DFG, German Research Foundation) under 
		Germany's Excellence Strategy - EXC 2121 ``Quantum Universe'' - 390833306.
		JSB acknowledges support by the Deutsche Forschungsgemeinschaft (DFG, German Research Foundation) through the CRC-TR 211 'Strong-interaction matter under extreme conditions', project number 315477589--TRR 211.
	\end{acknowledgments}
	
	\bibliographystyle{apsrev4-2}
	\bibliography{neue_bib_22}
\end{document}